\documentclass[aps,prl,twocolumn,floats,showpacs,superscriptaddress]{revtex4}
\usepackage{graphicx,epsfig}
\usepackage{times}
\usepackage{graphics,dcolumn,bm,float}
\usepackage{amssymb,amsmath,rotate,color}

\usepackage{graphicx}

\begin{document}
\unitlength 1 cm
\newcommand{\be}{\begin{equation}}
\newcommand{\ee}{\end{equation}}
\newcommand{\bearr}{\begin{eqnarray}}
\newcommand{\eearr}{\end{eqnarray}}
\newcommand{\nn}{\nonumber}
\newcommand{\la}{\langle}
\newcommand{\ra}{\rangle}
\newcommand{\cd}{c^\dagger}
\newcommand{\vd}{v^\dagger}
\newcommand{\ad}{a^\dagger}
\newcommand{\bd}{b^\dagger}
\newcommand{\kt}{{\tilde{k}}}
\newcommand{\pt}{{\tilde{p}}}
\newcommand{\qt}{{\tilde{q}}}
\newcommand{\eps}{\varepsilon}
\newcommand{\bpp}{\vec {p'}}
\newcommand{\bk}{{\vec k}}
\newcommand{\bp}{{\vec p}}
\newcommand{\bq}{{\vec q}}
\newcommand{\br}{{\vec r}}
\newcommand{\bR}{{\vec R}}
\newcommand{\up}{\uparrow}
\newcommand{\down}{\downarrow}
\newcommand{\fns}{\footnotesize}
\newcommand{\ns}{\normalsize}
\newcommand{\cdag}{c^{\dagger}}
\newcommand{\nup}{\nu_+}

\title{Nonlinear optical response in gapped graphene}

\author{S. A. Jafari{\footnote {Electronic address: akbar.jafari@gmail.com}}}
\affiliation{Department of Physics, Sharif University of Technology, Tehran 11155-9161, Iran}
\affiliation{School of Physics, Institute for Research in Fundamental Sciences, Tehran 19395-5531, Iran}

\begin{abstract}
We present a formulation for the nonlinear optical response in gapped graphene, where the
low-energy single-particle spectrum is modeled by massive Dirac theory. As a representative 
example of the formulation presented here, we obtain closed form formula for the
third harmonic generation (THG) in gapped graphene. It turns out that the covariant 
form of the low-energy theory gives rise to a peculiar logarithmic singularities in the 
nonlinear optical spectra. The universal functional dependence of the response function
on dimension-less quantities indicates that the optical nonlinearity
can be largely enhanced by tuning the gap to smaller values.
\end{abstract}
\maketitle

\section{Introduction}
Graphene is the first example of the realization of a truly two-dimensional (2D) crystal 
made of carbon atoms~\cite{Novoselov}. This intriguing material in addition to promise
for novel applications, from a fundamental point of view, provides 
the condensed matter community with a low-energy laboratory of 
Dirac electrons on the table top~\cite{NetoRMP}. Since then there has been 
remarkable success in pushing the idea of employing electronic, mechanical 
and various other properties of graphene in technological applications~\cite{Lin,Geim}.
The chiral nature of Dirac charge careers in graphene makes them
non-stoppable which mathematically manifests as the absence of
back-scattering. This means that with pristine graphene which 
contains massless Dirac fermions, there can be no "off" state in
electronic applications. Therefore, to enable the use of graphene 
in electronic devices, one needs to open up a gap in the single 
particle spectrum~\cite{Han} usually by statically reducing the symmetry
via extrinsic effects~\cite{Lanzara},
or by coupling to another field to generate dynamical masses~\cite{Kibis}.
Massive Dirac fermions possess a single-particle gap which 
causes the graphene to behave like a truly 2D semiconductor in some
respects. Nevertheless the nature of such "relativistic" massive theory is 
drastically different from the usual parabolic bands in a semi-conductor. 

Before the synthesis of graphene, the problem of two-dimensions
has been usually approached from the third dimension by e.g. a geometrical
confinement in hetero-structures~\cite{Bastard}, or by appropriately chosen 
B-field which effectively confines the dynamics of carriers into two
spatial dimensions~\cite{Koch}. In this respect, truly 2D 
gapped graphene provides a novel platform for non-linear optical applications.
The optical properties of 2D graphene have been extensively investigated at 
linear level~\cite{Mak, Pedersen}. However, at nonlinear level, there has been
a limited number of studies: A classical theory 
for the electromagnetic (EM) response of pristine graphene was developed
by Mikhailov and Zeigler~\cite{Mikhailov} who attributed strong nonlinear
EM response to the massless nature of Dirac electrons in graphene.
Nonlinear current response of massless Dirac fermions was studied by 
Wright and coworkers, who used direct expansion of the wave function 
in terms of multiples of frequency of the applied electric field~\cite{Wright}.
They found a high triple-frequency current response in massless graphene.
On the experimental side, Hendry and collaborators used four-wave mixing
to study the nonlinear optical response in graphene flaks, where they found
a remarkable large third-order optical response~\cite{Hendry}.
Also broadband optical nonlinearity was observed by Wang and 
coworkers in graphene dispersions~\cite{Wang}.

Therefore it is timely to consider the problem of third order
optical response in the 2D lattice of graphene by generalizing
it to the massive case. The problem of third order optical response 
for 1+1 dimensional massive Dirac fermions has been previously considered 
by Wu~\cite{Wu}. So the present work can also be considered as a natural 
generalization of the Wu's work to 2+1 dimensions. In one spatial
dimension, or for quasi one-dimensional systems such as organic materials
or carbon nano-tubes closed form expressions for the optical 
non-linear responses can be found~\cite{Margulis,Zarifi}. We employ a 
formulation we have developed earlier for investigation of the nonlinear 
optical properties in the matrix form~\cite{JafariSDW}. We find
that the covariant form of the dispersion relation for massive
Dirac fermions in 2+1 also allows for a simple closed form expression
for the third order response in general. We develop the general theory 
theory of nonlinear optical response for 2+1 Dirac fermions with arbitrary
gap parameter in terms of the Feynman diagrams. Then as a representative
example, we evaluate the line-shape of the third harmonic generation (THG).
We obtain a universal logarithmic functional form which depends only on the
combination $\nu/m$, where $\nu$ is the photon energy, and $m$ is the gap 
parameter.


\section{Model and method}
The single-particle energy band structure of Graphene consists of two 
Dirac cones which are connected to each other
by time-reversal symmetry. For optical applications the momentum of light
compared can be safely ignored and hence all optical processes of interest
will take place around a single cone. Therefore it is sufficient to consider
only one valley corresponding to which a spinor 
$\psi^\dagger_{\bp}=(a^\dagger_{\bp},b^\dagger_{\bp})$ denotes creation of 
$p_z$ electrons of momentum $\bp$ in sub-lattices A, B, respectively.
Then the effective low-energy Hamiltonian around the valley under consideration
can be written as,
\be
   H=\sum_\bp \psi^\dagger_\bp v_F (\bp.{\vec \sigma} + mv_F\sigma_z) \psi_\bp
\ee
where $\bp=p_x+i p_y\equiv p~ e^{i\varphi_\bp}$ is the complex representation
of a 2D vector. For simplicity we work in units with $v_F=e=\hbar=1$. 
The physical constants can be restored at the end of calculations if required~\cite{Desloge}.
The above Hamiltonian can be brought into
diagonal form if the following unitary transformation is applied 
\be
   V_\bp=\frac{1}{\sqrt{2\eps_\bp}} 
   \left(\begin{array}{cc}
       \frac{pe^{-i\varphi_\bp}}{\sqrt{\eps_\bp-m}} & \frac{pe^{-i\varphi_\bp}}{\sqrt{\eps_\bp+m}}\\
       \sqrt{\eps_\bp-m} & -\sqrt{\eps_\bp+m}
   \end{array}\right),~\eps_\bp=\sqrt{p^2+m^2}
\ee
to induce the change of basis,
\be
   \left(\begin{array}{c}
       a_\bp \\ b_\bp
   \end{array}\right)
   = V_\bp \left(\begin{array}{c}
       c_\bp \\ v_\bp
   \end{array}\right) 
   \leftrightarrow
   \psi_\bp = V_\bp \phi_\bp
\ee
to a new basis of conduction ($c_\bp$) and valence ($v_\bp$) bands, $\phi_\bp=(c_\bp,v_\bp)$,
with correspondingly positive and negative energies $\epsilon_\bp^\pm=\pm\eps_\bp$.
The diagonal form of the Hamiltonian in the new basis, $\phi_\bp$ reads,
\be
   H=\sum_\bp \phi^\dagger_\bp \eps_\bp \sigma_z \phi_\bp.
\ee

To proceed further, we need the explicit expression for the current operator
of 2+1 dimensional Dirac electrons in the new basis. 
Therefore we derive in the following the current
operator as follows: The current operator in terms of original real space
spinor $\psi_\bp$ can be written as,
\bearr
   {\vec J} &=&\sum_\bp \psi^\dagger_\bp \left[\frac{\partial}{\partial \bp} 
   (\bp.{\vec\sigma}+m\sigma_z)\right] \psi_\bp 
   =\sum_\bp \psi^\dagger_\bp {\vec \sigma} \psi_\bp
\eearr
The transformation $\psi_\bp = V_\bp \phi_\bp$ on the spinors, gives
the following form,
\be
   J_i=\sum_\bp \phi^\dagger_\bp V_\bp^\dagger \sigma_i V_\bp \phi_\bp,
   ~~~~~i=x,y  \label{currentop.eqn}
\ee
where the transformed Pauli matrices are given by,
\bearr
   V^\dagger_\bp \sigma_x V_\bp &=& \frac{-m\cos\varphi_\bp}{\eps_\bp}\sigma_x
   +\sin\varphi_\bp ~\sigma_y+\frac{p\cos\varphi_\bp}{\eps_\bp}\sigma_z\\
   V^\dagger_\bp \sigma_y V_\bp &=& \frac{-m\sin\varphi_\bp}{\eps_\bp}\sigma_x
   -\cos\varphi_\bp ~\sigma_y+\frac{p\sin\varphi_\bp}{\eps_\bp}\sigma_z
\eearr
Note that the second equation can be obtained from the first one by 
simply replacing $\varphi\to \varphi-\pi/2$, which is a consequence
of the vector character of the three Pauli matrices under SO(3) rotations.

\section{Multi-current correlations}
As detailed in Ref.~\cite{JafariSDW}, we are interested in calculation of
multi-current loops to which photon propagators are attached. As a demonstration
of this method, in the following we first calculate the two-current response
which is connected to optical conductivity. After reproducing well-known results,
along the same lines we proceed to calculate four-current correlations within
our matrix diagrammatic formulation which is suitable for situations
such as massive Dirac fermions in gapped graphene.

\begin{figure}[h]
  \begin{center}
  \includegraphics[width=6.0cm]{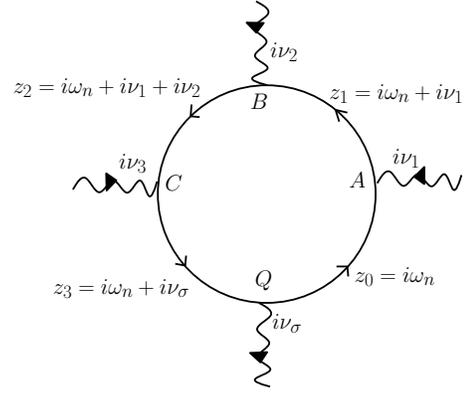}
  \caption{ 
  A typical Feynman diagram corresponding to a four-operator correlation
  function $\langle ABCQ\rangle$ contributing in the nonlinear optical response.
  Operators $A,B,C,Q$ could be any matter operator that couples
  to some power of the gauge field $\vec A$ of the incident light denoted
  by wavy lines (photons). The frequency $\nu_\sigma$ is the sum of
  all incoming frequencies $\nu_1+\nu_2+\nu_3$.
  }
  \label{loop.fig}
  \end{center}
\end{figure}

In general the fully retarded expectation value of nested commutators
of various operators, e.g. $A, B, C, Q$ schematically depicted in Fig.~\ref{loop.fig}
can be Lehman represented as~\cite{JafariSDW,JafariOC},
\be
   2\pi\delta(\nu+\nu_\sigma) \sum_{abc}\sum_{\cal P}
   \frac{A_{0a} B_{ab}C_{bc}Q_{c0}}{(\nu_1-E_{0a})(\nu_2-E_{ab})(-\nu_\sigma+E_{0c})}
\ee
where the sum of frequencies is $\nu_\sigma=\nu_1+\nu_2+\nu_3$, and  
$\sum_{\cal P}$ permutes $(A,\nu_1),(B,\nu_2),(C,\nu_3),(Q,-\nu_\sigma)$ 
around the current loop, and $B_{ab}$, etc. stand for the matrix element
$\langle a | B | b\rangle$. Note that in the above formula, the substitution
$\nu\to\nup=\nu+i0$ for all frequencies is understood.

\subsection{Linear response: Optical conductivity}
Within this formulation, the two-current response is given by~\cite{JafariSDW},
\begin{widetext}
\be
   \langle J_x J_x \rangle = 
   Tr \frac{1}{\beta}\frac{1}{L^2}\sum_{\cal P}\sum_\bp\sum_{i\omega_n}
   \left[ \frac{-m\cos\varphi_\bp}{\eps_\bp}\sigma_x +\sin\varphi_\bp ~\sigma_y \right]
   \frac{z_0+\eps_\bp\sigma_z}{z_0^2-\eps_\bp^2}
   \left[ \frac{-m\cos\varphi_\bp}{\eps_\bp}\sigma_x +\sin\varphi_\bp ~\sigma_y \right]
   \frac{z_1+\eps_\bp\sigma_z}{z_1^2-\eps_\bp^2}
\ee
\end{widetext}
where $i\nu$ is the external (photon) frequency. The corresponding diagram  is simpler than 
the one in Fig.~\ref{loop.fig}, where only two photons are attached, and only two
frequencies $z_{\alpha}=i\omega_n+i\alpha\nu,~\alpha=0,1$ are present. 
In the above equation, quantities in brackets 
are inter-band terms of the current operator and contain all necessary information
about the matrix elements of the current operator between the valence and 
conduction band. The matrix multiplications required in the above expression 
are easily performed to simplify the result as,
\be
   \langle J_x J_x \rangle = 2! Tr \frac{1}{\beta}\frac{1}{L^2}\sum_\bp\sum_{i\omega_n}
   \frac{(z_0 z_1-\eps_\bp^2)(m^2+p^2\sin^2\varphi_\bp^2)}
   {(z_0^2-\eps_\bp^2) (z_1^2-\eps_\bp^2) \eps_\bp^2}.
\ee
Note that the sum over permutations of two $J_x$ operators
around the loop produces a $2!$ factor in the right hand side~\cite{JafariSDW},
as the above formula is symmetric with respect to fermion propagation frequencies $z_0,z_1$.
Standard Matsubara summation, the two-current correlation function 
simplifies to,
\be
   \int\frac{2d^2\bp}{(2\pi)^2}
   \left[\frac{1}{i\nu-2\eps_\bp}-\frac{1}{i\nu+2\eps_\bp}\right]
   \times \frac{m^2+p^2\sin^2\varphi_\bp}{\eps_\bp^2}
\ee
where the temperature is assumed to be zero so that the Fermi function
is $1$ only for negative energies, and is zero otherwise. 
After analytic continuation, $i\nu\to \nu+i\eta$, the spectral
function (imaginary part) corresponding to the absorption -- i.e. 
the first term in the bracket -- can be easily obtained. To compare
with the existing results in the literature, we use the conductivity formula
$\sigma_{xx}(\nu)=e^2\langle J_xJ_x\rangle/(i\nu)$ and restore the 
fundamental constants to obtain
\be
   \mbox{Re}\sigma_{xx}(\nu)=\frac{e^2}{8\hbar}
   \left[1+\frac{4m^2}{\nu^2} \right]
   = \frac{\sigma_0}{2} \left[1+\frac{4m^2}{\nu^2} \right]\label{optcond.eqn}
\ee
which after noting the fact that our calculation has been 
done for a single-valley, agrees with e.g. eq. (10) of Ref.~\cite{Kotov}.
Note that here $\sigma_0=e^2/(4\hbar)$ is the conductance of
ideal graphene.

In a similar way, one can calculate correlation functions like 
$\langle J_xJ_y\rangle$. In simplifying such elements one should
note that odd powers of $\sin\varphi$ or $\cos\varphi$ average to zero
upon $\varphi$ integration. Moreover, summation over permutations $\cal P$
picks up the most symmetric part. In the case of two-current correlation
the $xy$ component will become zero, unless a magnetic field is applied~\cite{Gusynin}.

\subsection{Four-current correlations}
Now that we checked our formulation with well-known results, let us
calculate higher order responses. But before 
proceeding to question of higher order responses, we note
that in a general material the higher order correlation between the
current operators, i.e. $\langle J_a J_b J_c J_d\rangle$ which essentially 
contains four velocity vertices along $a,b,c,d\in\{x,y\}$ directions of space,
is not the only important term when one is interested in third order
response. In general terms of the type $J_a \tau_{bc} J_d$, etc also
may appear where the stress tensor $\tau_{bc}$ too contributes 
to the third order response~\cite{JafariOC}. Similar to current vertices 
given by 
\be
   v_a = \frac{\partial \eps_\bp}{\partial p_a}
\ee
even-parity vertices of the form
\be
   \frac{\partial \eps_\bp}{\partial p_c \partial p_d}
\ee
may also enter the theory of nonlinear optical response for materials
with arbitrary band structure defined by $\eps_\bp$. 
In the case of ideal graphene ($m=0$), it is interesting to note that, due to the
linear form of energy as a function of momentum $\bp$, only velocity
vertices will appear in the theory as we calculated them in \eqref{currentop.eqn}.
All higher order derivatives related to the stress tensor, etc. will be 
identically zero for massless Dirac  fermions. In the case of massive
Dirac fermions where the asymptotic behavior of the energy dispersion
for photon energy scales much higher than the gap parameter $m$,
becomes linear, the terms arising from stress tensor will have negligible
contributions, for energy scales much beyond those corresponding to
band edge excitations.

\begin{figure}[tb]
  \begin{center}
  \includegraphics[width=4.0cm]{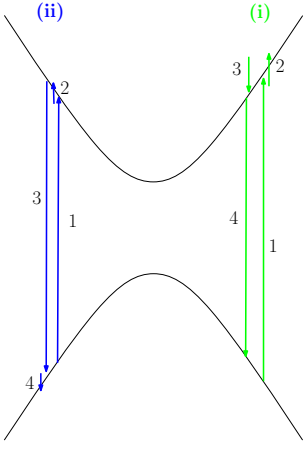}
  \caption{(Color online) Schematic sketch of two types of operator
  sequences contributing to the four-operator correlation function.
  Type (i) and (ii) processes are denoted by green and blue, respectively.
  Indices $1$ to $4$ label the operators from right to left.
  }
  \label{i-ii.fig}
  \end{center}
\end{figure}

With this point in mind, in this paper we focus on
four-current correlations of the form $\langle J_xJ_xJ_xJ_x\rangle$. 
From the rightmost current operator, 
only the inter-band terms contribute which lead to an excitation into
the conduction band. Then the second operator from right will have to
create intra-band excitation.
However, for the third and fourth operator,
there are two different possibilities depicted in
Fig.~\ref{i-ii.fig}. Either the operator number $3$ creates an intra-band
excitation as in part (i) of Fig.~\ref{i-ii.fig}. In this case the operator 
number $4$ must create an inter-band excitation. Such sequence of 
operators can be represented by $-++-$, where $-$ denotes {\em inter}-band vertex, 
and $+$ stands for an {\em intra}-band vertex. With this notation, 
the second possibility denoted in part (ii) of Fig.~\ref{i-ii.fig}
can be summarized as $+-+-$, i.e. the third operator returns the system
to valence band, and operator number $4$ has to create an intra-band
excitation. Note that a sequence like $----$ is not a genuine four-operator
correlation, but rather a product of two-operator correlations and
can be extracted from linear responses as well. 
All other possibilities can be generated by the sum over 
permutation $\sum_{\cal P}$. Therefore the above sequences
are two general families of excitation patterns which by the cyclic 
property of the trace involved in the {\em loop} diagram of Fig.~\ref{loop.fig}, 
include all possibilities. These are summarized in Fig.~\ref{i-ii.fig}.

Now let us calculate the contribution of two sequences of 
excitations (i) and (ii). First note that the only terms surviving 
the trace operation are those proportional to unit $2\times 2$ matrix, and that odd 
powers of $\sin\varphi$ or $\cos\varphi$ do not survive the angular 
integration arising from $\sum_\bp$. Therefore, the non-zero terms 
of the first class of terms can be simplified after some algebra as,
\bearr
   &&2\sum_{\cal P}\frac{1}{\beta}\sum_{i\omega_n} \int\frac{d^2\bp}{(2\pi)^2}
   \left[\prod_{\alpha=0}^3\frac{1}{z_\alpha^2-\eps_\bp^2}\right] \times \\
   &&\frac{p^2\cos^2\varphi_\bp}{\eps_\bp^2}. 
   \left[\frac{m^2\cos^2\varphi_\bp}{\eps_\bp^2}+\sin^2\varphi_\bp\right]\times \nn\\
   &&\left[(z_1z_2+\eps_\bp^2)(z_0z_3-\eps_\bp^2)+\eps_\bp^2(z_1+z_2)(z_0-z_3) \right],\nn
\eearr
where frequencies $z_\alpha,\alpha=0,1,2,3$ are defined in Fig.~\ref{loop.fig}.
Similarly for the class (ii) terms we obtain,
\bearr
   &&2\sum_{\cal P}\frac{1}{\beta}\sum_{i\omega_n} \int\frac{d^2\bp}{(2\pi)^2}
   \left[\prod_{\alpha=0}^3\frac{1}{z_\alpha^2-\eps_\bp^2}\right] \times \\
   &&\frac{p^2\cos^2\varphi_\bp}{\eps_\bp^2}. 
   \left[\frac{m^2\cos^2\varphi_\bp}{\eps_\bp^2}+\sin^2\varphi_\bp\right]
   \times \nn\\
   &&\left[ 
   \eps_\bp^2(z_2-z_3)(z_0-z_1)-(\eps_\bp^2-z_2z_3)(\eps_\bp^2-z_0z_1)
   \right],\nn
\eearr
where an overall factor of $2$ arises from trace over the unit matrix.
Adding the above terms gives,
\bearr
   &&2\sum_{\cal P}\frac{1}{\beta}\sum_{i\omega_n} \int\frac{d^2\bp}{(2\pi)^2}
   \left[\prod_{\alpha=0}^3\frac{1}{z_\alpha^2-\eps_\bp^2}\right] \times \nn \\
   &&\frac{p^2\cos^2\varphi_\bp}{\eps_\bp^2}. 
   \left[\frac{m^2\cos^2\varphi_\bp}{\eps_\bp^2}+\sin^2\varphi_\bp\right]\times 2\eps_\bp^2 \times\nn\\
   && \left[z_0z_1+z_0z_2-z_1z_2-\eps_\bp^2 \right].
\eearr
Now we can evaluate the Matsubara sums, and at the end the result
will be symmetrized with respect to exchange of the frequencies of
the attached photon vertices. Using the angular averages,
\bearr
   \langle \cos^4\varphi \rangle =\frac{6}{8},~~~~~
   \langle \cos^2\varphi~\sin^2\varphi \rangle=\frac{1}{8},
\eearr
the angular integration can be simplified to,
\bearr
   &&\sum_{\cal P}\frac{1}{\beta}\sum_{i\omega_n} \int\frac{pdp}{2\pi}
   \left[\prod_{\alpha=0}^3\frac{1}{z_\alpha^2-\eps_\bp^2}\right] \times\nn \\
   &&\frac{p^2}{2} \left[6\frac{m^2}{\eps_\bp^2}+1\right] 
   \left[z_0z_1+z_0z_2-z_1z_2-\eps_\bp^2\right].\label{chi3.eqn}
\eearr
Up to this point, the formulation is quite general. At this point,
let us specialize to the case of THG, where
$i\nu_1=i\nu_2=i\nu_3=i\nu$ and $i\nu_\sigma=3i\nu$. In this case,
the Feynman diagram shown in Fig.~\ref{loop.fig} must be labeled with 
$z_\alpha=i\omega_n+\alpha i\nu$ for $\alpha=0,1,2,3$. 
Permuting the external vertices amounts to moving the outgoing
$3i\nu$ photon frequency around the loop. Such permutation can
be accounted for by a cyclic permutation of the set $\{z_0,z_1,z_2,z_3\}$.
Performing the $\sum_{\cal P}$ will only affect the last term in 
Eq.~\eqref{chi3.eqn} as,
\be
   \sum_{\cal P} \left[z_0z_1+z_0z_2-z_1z_2-\eps_\bp^2\right]
   =2\left[z_0z_2+z_1z_3-2\eps_\bp^2 \right], \nn
\ee
where the THG identification, $z_\alpha=i\omega_n+\alpha i\nu$ with $\alpha=0,1,2,3$
is understood. Let us emphasize that the above result for the sum
over permutation holds only for THG. Therefore the THG susceptibility
will become,
\bearr
   \chi^{\rm THG}&=&\frac{1}{\beta}\sum_{i\omega_n} \int\frac{pdp}{2\pi}
   \left[\prod_{\alpha=0}^3\frac{1}{z_\alpha^2-\eps_\bp^2}\right] \times\\
   && p^2 \left[6\frac{m^2}{\eps_\bp^2}+1\right] 
   \left[z_0z_2+z_1z_3-2\eps_\bp^2\right]\nn
\eearr
The $1/\beta\sum_{i\omega_n}$ can be performed with standard contour
integration techniques producing Fermi functions $f(\pm\eps_\bp)$
corresponding to poles at conduction and valence bands, respectively.
Assuming the temperature to be zero, and for undoped massive Dirac
spectrum, only contributions from poles at the valence band will be 
left for which the residues can be calculated to give,
\bearr
   \chi^{\rm THG}(\nu_+)&=& \int\frac{pdp}{2\pi}
   p^2 \left[6\frac{m^2}{\eps_\bp^2}+1\right] \times \label{thg1.eqn}\\
   &&\frac{3}{2}\frac{4\eps_\bp^2+\nu_+^2}{\eps_\bp
   (\nu_+^2-\eps_\bp^2)(\nu_+^2-4\eps_\bp^2)(\nu_+^2-9\eps_\bp^2)}\nn
\eearr
Note that the analytic continuation $i\nu\to\nu_+=\nu+i0$ 
has been performed in the above equation~\cite{JafariSDW}.

\begin{figure}[t]
  \begin{center}
  \includegraphics[width=8.8cm]{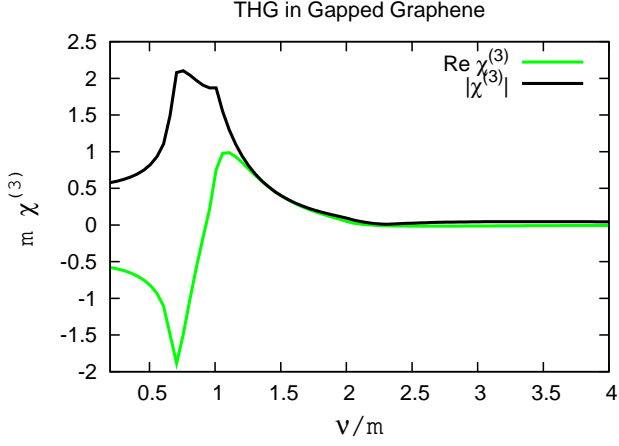}
  \caption{(Color online) THG line-shape for massive Dirac fermions in gapped graphene. The
  gap parameter $m$ is taken to be unit of energy in this figure. Fundamental constants
  are assumed to be $e=\hbar=1$. }
  \label{lineshape.fig}
  \end{center}
\end{figure}

\section{Result and discussions}
The radial momentum integration in Eq.~\eqref{thg1.eqn} can be performed with the
change of variable to $\varepsilon^2=p^2+m^2$ which gives the
following closed form result
\bearr
   &&\chi^{\rm THG}(\nup)=-\frac{m^3}{2\pi\nup^4}+\frac{1}{192\pi\nup^5}\times  \nn\\
   &&\left[-24(6m^2+\nup^2) (m^2-\nup^2 )\ln\frac{m+\nup}{m-\nup} \right. \nn\\
   &&+3(4m^2-\nup^2)(24m^2+\nup^2)\ln\frac{2m+\nup}{2m-\nup} \nn\\
   &&\left.+(4m^2-9\nup^2)(8m^2+3\nup^2)\ln\frac{2m+3\nup}{2m-3\nup}\right]
   \label{thg2.eqn}
\eearr
First of all, note that the above expression when the fundamental constants
are restored will be proportional to $e^4/\hbar^3$. Therefore the above
expression shows the results in the natural units. A remarkable feature 
of the above expression is that, the quantity $m\chi^{\rm THG}$ is only 
a function of $\nu/m$. This functional form is universal characteristic
of 2+1 dimensional Dirac fermions. Hence in Fig.~\ref{lineshape.fig}, we have
plotted the real part and the magnitude of THG line-shape incorporating
this observation. This peculiar scaling form indicates that in the limit
where $m\to 0$, and the system approaches the gapless limit, since 
$m\chi^{\rm THG}$ in the natural units employed here will have to remain 
on the scale of unity, the $\chi^{\rm THG}$ itself will grow inversely
proportional to the gap parameter $m$. This sheds a new light into why
pristine (gapless) graphene is expected to display large optical 
nonlinearity~\cite{Hendry}.

Apart from the zero frequency divergence, which is a general feature of 
optical response functions,
an interesting point in the above expression which can be noticed is that,
it is precisely the covariant (relativistic) form of the dispersion relation
$\varepsilon^2=p^2+m^2$, which after changing integration variables from
momentum $p$, to energy $\varepsilon$ leads to logarithmic dependence 
at finite frequencies $\nu_\ell/m=2/\ell$, where $\ell$ is an integer $\ell=1,2,3$.
This would not be the case for a typical parabolic dispersion relation
in ordinary semiconductors. 
The main THG peak corresponding to $\ell=1$ and the second one corresponding to $\ell=2$
are clearly seen in the THG spectrum depicted in Fig.~\ref{lineshape.fig}. 
The feature corresponding to $\ell=3$ is the weakest feature. 
Note that as we calculated in Eq.~\eqref{optcond.eqn},
such logarithmic dependence does not show up in the linear optical response.
Therefore it can be considered as a valuable information which is contained
only in higher order optical responses. Moreover, this logarithmic 
dependence is not restricted to the THG spectrum, but it also appears
in all other higher order responses. To see this, we note in 
Eq.~\eqref{chi3.eqn} that in the special case of THG, only the functional form 
of $\sum_{\cal P}$ as a function of loop frequencies $z_\alpha$, will be affected.
But the dimensional form of the summand will generally retain a similar form. 

The calculation of THG for massive Dirac fermions in 1+1 dimension by Wu~\cite{Wu}
indicates a characteristic inverse square root divergence at the main THG peak $\nu_{\ell=1}$
and its harmonics. In the case of small clusters (of correlated 2D nature), 
a numerical investigation of THG spectrum suggests a line-shape composed of
a superposition of resonances corresponding to simple poles~\cite{Takahashi}.  
For extended 2D system with broken symmetry, it was shown that the nesting underlying
spin density wave instability can produce a peculiar $z^{-1/2}\ln(z)$ form of 
singularity in the non-linear optical spectra~\cite{JafariSDW}. With the above
examples in mind, the gapped graphene as a first example of truly two dimensional system
provides us with a unique example of a 2D system where optical nonlinearity
manifests in a universal logarithmic function of the dimension-less quantity 
$\nu/m$.

It is interesting to check the limit of gapless graphene by taking 
the limit $m\to 0$ in Eq.~\eqref{thg2.eqn}, which gives,
\be
   m\chi^{\rm THG}(\nu)=\frac{54i}{192}\frac{m}{\nup}
\ee
The coefficient of the above expression is universal and can be
regarded as a hallmark of the ideal graphene at the THG spectroscopy.

\section{Acknowledgements}
This work was supported by the National Elite Foundation (NEF) of Iran.

\end{document}